# Shaping nanoscale magnetic domain memory in exchange-coupled ferromagnets by field cooling

Karine Chesnel[1], Alex Safsten[1], Matthew Rytting[1] & Eric E. Fullerton[2]

The advance of magnetic nanotechnologies relies on detailed understanding of nanoscale magnetic mechanisms in materials. Magnetic domain memory (MDM), that is, the tendency for magnetic domains to repeat the same pattern during field cycling, is important for magnetic recording technologies. Here we demonstrate MDM in [Co/Pd]/IrMn films, using coherent X-ray scattering. Under illumination, the magnetic domains in [Co/Pd] produce a speckle pattern, a unique fingerprint of their nanoscale configuration. We measure MDM by cross-correlating speckle patterns throughout magnetization processes. When cooled below its blocking temperature, the film exhibits up to 100% MDM, induced by exchange-coupling with the underlying IrMn layer. The degree of MDM drastically depends on cooling conditions. If the film is cooled under moderate fields, MDM is high throughout the entire magnetization loop. If the film is cooled under nearly saturating field, MDM vanishes, except at nucleation and saturation. Our findings show how to fully control the occurrence of MDM by field cooling.

[1] Department of Physics and Astronomy, Brigham Young University, Provo, Utah 84602, USA. [2] Center for Memory and Recording Research, University of California San Diego, La Jolla, California 92093-0401, USA. Correspondence and requests for materials should be addressed to K.C. (email: kchesnel@byu.edu).

Understanding the mechanisms driving nanoscale magnetic reversal and memory phenomena in layered magnetic films is essential for the development of magnetic recording and spintronic nanotechnologies[1–4]. Thin ferromagnetic (F) films with perpendicular magnetic anisotropy (PMA) are used for high-density magnetic recording applications because they allow the control of nanometric magnetic domains to encode information[5–9]. Interlaying PMA films with other F, antiferromagnetic (AF) or non-magnetic layers in nano-patterned composite structures has enabled the development of magnetic memory devices[10–12]. The layering of PMA films, especially exchange-coupling F layers with AF layers, leads to specific magnetic domain topologies and associated memory properties. These mechanisms are however very complex and still require more detailed understanding.

Although studies of PMA media are generally based on macroscopic magnetometry observations[13,14], a microscopic view of the magnetic properties, which can be obtained via magnetic microscopy[15–18] and X-ray scattering[19–22], is often lacking. We present here a unique coherent X-ray magnetic correlation study, mapping out nanoscale magnetic memory phenomena in exchange-biased PMA multilayers throughout their magnetization process. We show how magnetic domain memory (MDM), that is, the ability for the magnetic domain pattern in an F layer to retrieve its exact same domain topology after field cycling, can be turned on or off and spatially controlled by tuning the nature of the exchange bias with an AF template.

Single PMA F films alone do not normally exhibit MDM. If the film is relatively smooth, magnetic domains nucleate and propagate randomly throughout the material when an external magnetic field is cycled. Even though the net magnetization of the film is uniquely defined at a given applied field and following a given magnetization history, the nanoscale domain pattern is usually not uniquely defined. Although the average domain size remains the same, an unlimited number of different domain patterns can lead to the same net magnetic moment. Each time the F film is saturated and magnetization is cycled, the domain pattern evolves through a new domain configuration. Consequently, the films show little or no MDM when the external magnetic field is cycled.

If the film is somewhat rough, some MDM may occur, induced by defects[23]. In this instance, the observed defect-induced memory is only partial (between 10 and 50%) and is highest in the nucleation phase of the magnetization process[24]. The defects tend to pin the nucleating sites for the magnetic domains, but the domain propagation remains mostly random, thus leading to a loss of MDM during the rest of the magnetization process.

Here we show how up to 100% MDM can be achieved, even in very smooth F films, by exploiting exchange coupling with an AF template. MDM can be turned on by cooling the film below the blocking temperature $T_B$ and it can be turned off again by heating above $T_B$. Moreover, we show how the spatial occurrence of MDM at the nanoscale can be shaped throughout the magnetization process by finely adjusting the amount of net exchange bias, that is, the magnitude of the applied field while cooling. Our preliminary studies revealing the occurrence of high MDM induced by exchange coupling[25], its spatial dependency[26] and temperature dependency[27] were limited to the specific case of zero-field cooling (ZFC). In the present study, we show a complete mapping of MDM that investigates a whole range of field-cooling (FC) conditions, thus revealing surprising behaviours when the cooling field is pushed towards saturation. These results show that nanoscale spatial dependence of domain memory can be controlled by varying the amount of exchange bias imprinted in the material.

## Results

**Macroscopic magnetic properties**. Our PMA F films are based on well-studied [Co/Pd] multilayers[8,13] that exhibits serpentine-shaped magnetic domains that are ~150–200 nm in width. In the absence of external magnetic field, these domains usually form a nanoscale labyrinthine pattern, as seen on Fig. 1a. To incorporate magnetic exchange coupling into the films[28], the [Co/Pd] F layers are periodically interlayed with thin AF layers made of an IrMn alloy, thus forming a $[[Co(4\,\text{Å})/Pd(7\,\text{Å})]_{\times 12}/IrMn(24\,\text{Å})]_{\times 4}$ multilayer. Dipolar fields correlate the domains in adjacent F layers and exchange coupling occurs between the F and AF layers when the film is cooled below a blocking temperature $T_B \sim 300\,K$. FC this film below $T_B$ induces significant modifications in the magnetization loop, enhancing the coercive field and inducing a net bias (shifting of the loop), as shown in Fig. 1b. At 20 K, the observed net bias is ~200 Oe, in the direction opposite to the field applied during the cooling (here +10 kOe). The loop is widened by ~300 Oe at its narrowest region (coercive point) and by ~600 Oe at its widest region (gap between saturation and nucleation points increasing from 2,400 Oe at 300 K to 3,000 Oe at 20 K). As in ZFC conditions[25], the widening of the loop results from exchange couplings between the Co spins in the F layer with the interfacial uncompensated Mn spins in the IrMn AF layer. The biasing of the loop is caused by the net magnetic moment imprinted in the IrMn after FC. These magnetization measurements give macroscopic information about the film's net magnetization $M$ only. To probe the associated magnetic domains at the nanoscale and its evolution throughout the magnetization process at low temperature, we use soft X-ray scattering. With a wavelength in the nanometric range, soft X-rays provide a perfect tool to resolve nano-sized magnetic domains and their spatial topological changes under the application of an *in-situ* magnetic field.

**X-ray magnetic scattering patterns**. The [Co/Pd]/IrMn magnetic films were studied via coherent X-ray resonant magnetic scattering (CXRMS)[29] using synchrotron radiation. In this CXRMS experiments, the films were exposed to coherent X-ray light and produced scattering patterns detected downstream, as shown in Fig. 1c. The films were mounted on a cryogenic holder to allow the cooling of samples down to ~20 K with liquid helium. An *in-situ* magnetic field was applied perpendicular to the films via sets of conically shaped octupolar electromagnets[30]. After being demagnetized at high temperature, around 400 K, the films were cooled down to 20 K under a given external magnetic field $H_{FC}$. The experiment was repeated in various FC conditions. The magnitude of the cooling field $H_{FC}$ varied from zero (ZFC) up to 3,200 Oe, which is near the saturation point. In each FC state, once the temperature had stabilized near 20 K, the magnetic field was cycled and CXRMS speckle patterns were collected throughout the magnetization process.

When illuminated by coherent X-rays tuned to magnetic resonance energies, the magnetic domains in the film scatter the light and produce a magnetic scattering pattern that is speckled. As the film is relatively smooth and fully dense[31], it does not produce any significant charge scattering signal and the observed signal is essentially magnetic, as illustrated by the selection of as-collected scattering images in Fig. 1d–f. At nucleation (Fig. 1d), the sparse nucleated domains produce a faint disk-like scattering signal; in the coercive region (Fig. 1e), the fully propagated magnetic domains produce a bright ring-like scattering pattern; at saturation (Fig. 1f), domains have disappeared and the scattering signal vanishes, thus confirming that the observed scattering signal is essentially magnetic. To ensure that we only retain the magnetic signal in our analysis, we subtract the saturation scattering pattern (Fig. 1f) from all the other patterns collected throughout the magnetization loop, before extracting the speckle signal.

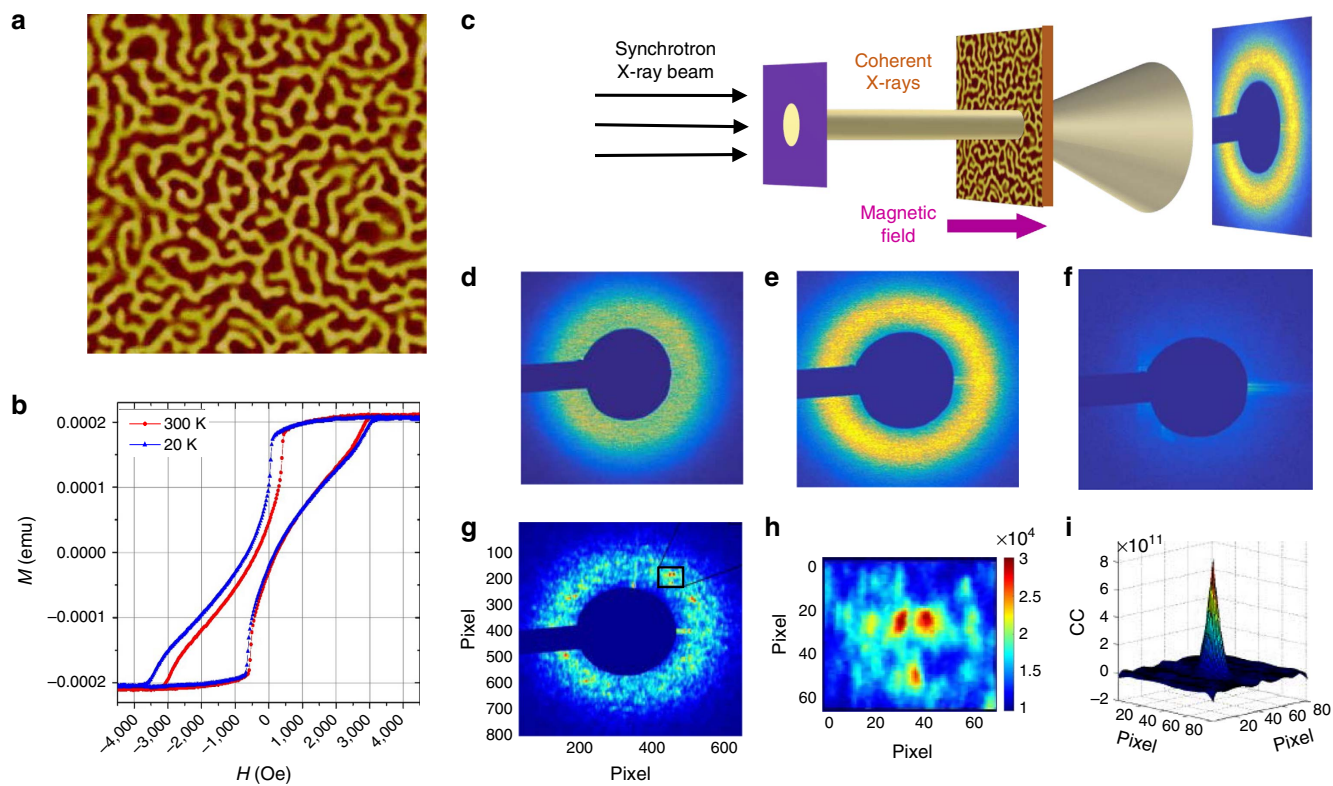

**Figure 1 | Imaging of magnetic domain memory in ([Co/Pd]/IrMn) via coherent X-ray scattering and speckle correlation.** (**a**) MFM image (5 × 5 μm$^2$) of the magnetic domains in the film at remanence at 300 K. (**b**) Magnetization loops $M(H)$ collected via VSM at 300 K and at 20 K after FC under a +10 kOe field, showing a biasing effect. (**c**) Layout of the CXRMS experiment. (**d**-**f**) CXRMS patterns as collected at the Co-$L_3$ edge (780 eV) on a field-cooled film at 20 K and at different field values on the ascending branch of the magnetization curve: (**d**) near nucleation ($H = -200$ Oe), (**e**) near the coercive point ($H = +500$ Oe) and (**f**) at saturation ($H = +4,000$ Oe). (**g**) Magnetic speckle pattern obtained after extracting the coherent scattering signal and removing the incoherent scattering signal from a CXRMS pattern. (**h**) Zoomed-in view of the speckle spots seen in the selected area of (**g**). The colour bar shows actual number of counts on the detector. (**i**) Typical correlation pattern $A \times B$ obtained by cross-correlating two speckle patterns $A$ and $B$ such as that in **g**. The $z$ axis, denoted CC, represents the cross-correlation value. The unit for CC is number of counts squared. The amount of correlation is evaluated by measuring the area under the peak and normalizing it with autocorrelation signals, resulting in the normalized correlation coefficient $\rho$.

The extracted speckle pattern shown in Fig. 1g was collected at the remanent point, where the magnetic domains form a maze such as that in Fig. 1a. Because of the labyrinthine nature of the domain pattern, the scattering pattern has an isotropic ring-like shape. The radius of the ring is inversely proportional to the average magnetic period in the domains pattern. In the coercive region, where the ring radius is the largest, that is, the magnetic domain size is the smallest, the measured magnetic period is around 340 nm, corresponding to an average width for the individual domains of around 170 nm. The coherence of the light produces interferences in the scattering pattern, leading to a collection of speckle spots, as seen in Fig. 1h. The speckle pattern, with its specific distribution, shape and location of the individual speckle spots, provide a unique fingerprint of the actual nanoscale magnetic domain pattern in the film.

**Magnetic correlation maps.** We probed MDM by cross-correlating speckle patterns collected at different field values throughout the entire magnetization cycle[32]. Speckle patterns were either from the same magnetic hysteresis cycle or from two separate cycles, the number of separating cycles being tracked. The amount of MDM was quantified by integrating the signal under the peak observed in correlation patterns as shown in Fig. 1i and a normalized correlation coefficient $\rho$, defined in the Methods section, was evaluated.

By cross-correlating thousands of speckle patterns collected throughout many cycles of the magnetization loop, we generated correlation maps, such as the ones shown in Figs 2–4. Each map plots the correlation coefficient $\rho$ (red being high and blue low on the colour scale) as a function of applied field. The coordinates on the maps are the field values ($H_1$ and $H_2$) of the two correlated speckle patterns. A vertical (or horizontal) slice through the map corresponds to a fixed field $H_1$ (or $H_2$) and a diagonal slices corresponds to $H_1 = H_2$, also known as return point memory[33]. Correlation maps were measured both on the ascending and the descending branches of the magnetization loop, within the same cycle or after certain number of cycles (up to 6 cycles). The collection of data on several subsequent cycles allowed us to check the reproducibility of the results and to take statistical averages. The results were highly reproducible. Thus, each correlation map shown here represents the correlation $\rho(H_1, H_2)$ value averaged over all the possible pairs of correlated cycles, for a given cycle separation.

**Magnetic correlations under moderate FC.** Correlation maps measured under moderate FC conditions are shown in Fig. 2. The associated cooling field was $H_{FC} = 1,280$ Oe on the ascending branch, right in the reversible region of the magnetization loop

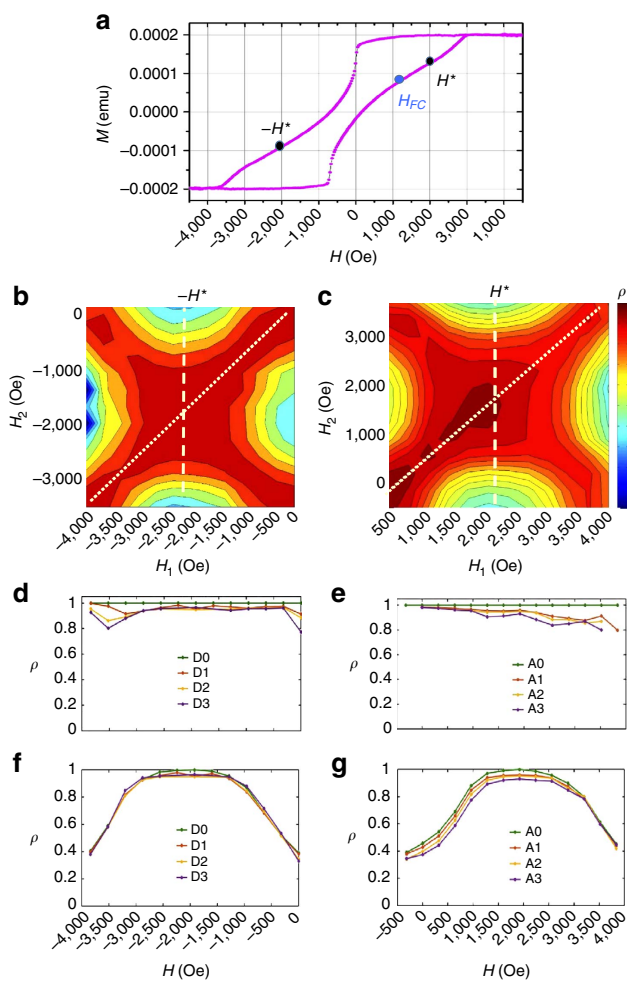

**Figure 2 | Correlations maps and slices measured at 20 K after FC under $H_{FC} = 1,280$ Oe.** (**a**) Magnetization curve $M(H)$ at 20 K after FC under $H_{FC} = +1,280$ Oe, a field point indicated by the blue dot on the curve. (**b,c**) Correlation maps $\rho(H_1, H_2)$ for one cycle separation (the correlation coefficient $\rho$ is plotted in colour, where red is high and blue is low) measured along (**b**) the descending branch and (**c**) the ascending branch. (**d,e**) Diagonal slices, for which $H_1 = H_2$ (dotted line). (**f,g**) Vertical slices through the maps at $H_1 = H^*$ (dashed line). These graphs show slices through successive correlation maps, using the notation A0, A1… D0, D1 and so on. In this notation, the letter A stands for the ascending branch, the letter D stands for the descending branch and the index corresponds to the number of cycles separating the two correlated images. For example, D3 shows correlation between two descending branches that are separated by three full cycles. In each map/slice, the plotted coefficient $\rho$ is an average value over many cycles at a fixed cycle separation.

(Fig. 2a). The resulting correlation maps for the descending branch (Fig. 2b) and for the ascending branch (Fig. 2c) show a very high correlation degree $\rho$ approaching 100% in the central region. The high MDM extends over a wide plateau across the map, centred about $H_1 = H_2 = H^* \sim 2,000$ Oe. This result persists throughout field cycling. Even after many cycles, the map is unchanged, as slices in Fig. 2d–g indicate. On the diagonal slices ($H_1 = H_2$), $\rho$ is close to 100% over the entire range of field. On the vertical slices at $H_1 = H^*$, $\rho$ reaches nearly 100% over a wide plateau extending from $\sim 1,000$ to 3,000 Oe. In that part of the magnetization loop, the magnetic domain pattern remains topologically close to the one at $H^*$. When the diagonal slice is performed on the same cycle maps, it is expected that $\rho = 100\%$, as, by definition, speckle patterns are correlated with themselves, but we find here that diagonally slicing subsequent maps also lead to $\rho \sim 100\%$. This suggests that during the magnetization process, magnetic domains nucleate, propagate and collapse following the same morphological path, guided by the template imprinted in the AF layer. This result using a moderate $H_{FC} = 1,280$ Oe cooling field agrees very well with results obtained in ZFC conditions when $H_{FC} = 0$. Similar to $H_{FC} = 0$, a moderate cooling field $H_{FC} = 1,280$ Oe leads to high MDM approaching 100% over a large field range[27] and also shows persistence through field cycling[34].

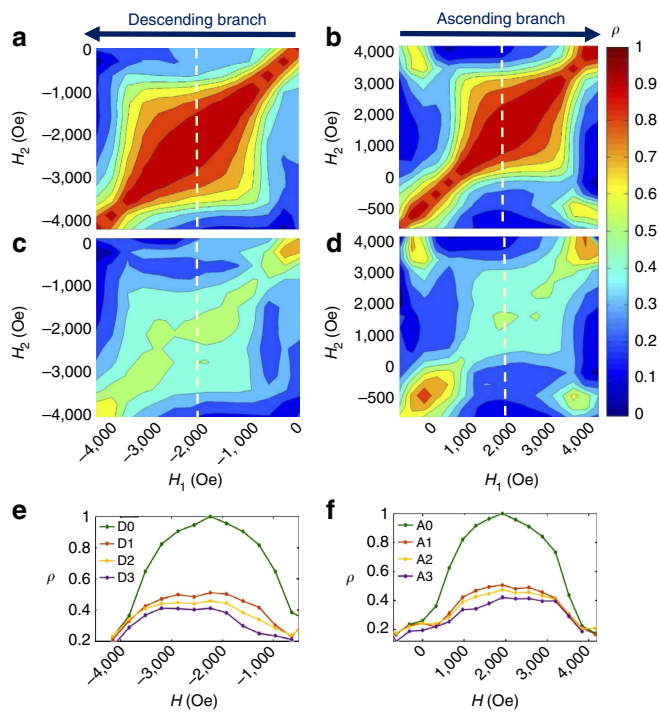

**Figure 3 | Correlations maps and slices measured at 20 K after FC under $H_{FC} = 3,200$ Oe.** (**a,b**) Correlation maps $\rho(H_1, H_2)$ measured within the same cycle on the descending and ascending branches, respectively. (**c,d**) Correlation maps $\rho(H_1, H_2)$ measured at one cycle separation. (**e,f**) Vertical slices at $H_1 = H^*$ (along the dashed line). These graphs show slices through successive correlation maps, using the notation A0, A1… D0, D1 and so on. In this notation, the letter A stands for the ascending branch, the letter D stands for the descending branch and the index corresponds to the number of cycles separating the two correlated images.

**Magnetic correlations under high FC.** A selection of correlation maps measured under high FC conditions is shown in Fig. 3. The associated cooling field was $H_{FC} = 3,200$ Oe on the ascending branch (with the cooling starting at 400 K), that is, near saturation. The resulting correlation maps show a clear loss of MDM in the central region of the magnetization loop. Correlation maps measured within the same cycle (Fig. 3a,b) still show some high correlation but not extending as much as for lower cooling field values. Here, the high correlation is concentrated on a narrow region along the diagonal ($H_1 = H_2$). In addition, as soon as one cycle has been completed, the correlation drastically drops to as low as 40% in the central region of the map (Fig. 3c,d). The loss of MDM occurs similarly on the ascending and descending branches. Both diagonal and vertical slices (Fig. 3e,f) through the maps clearly show the drastic drop in $\rho$, from 100% to $\sim 50\%$, after one cycle. During subsequent cycles, $\rho$ continues to gradually drop, from $\sim 50\%$ down to 40% after three cycles.

The maps show a higher correlation at the four corners, when $H_1$ and $H_2$ are both near one extremity, either nucleation ($H \sim -500$ Oe on ascending branch) or saturation ($H \sim 4,000$ Oe). The occurrence of a high correlation at all the four corners of the correlation map indicates that the magnetic domain pattern at saturation is correlated with the magnetic domain pattern at nucleation. This interesting correlation emerges when the cooling field is high, nearly saturating. In such cooling conditions, the domain pattern imprinted in the AF layer via the uncompensated spins is made of small isolated domains, sparsely scattered throughout the film. Owing to the exchange coupling, these sparse domains in the AF layer then tend to drive the topology of magnetic domains in the F layer when they are nucleating but also when they are approaching saturation. Therefore, the magnetic domain topology in the F layer at nucleation and at saturation both tend to match quite well the imprinted pattern. Consequently, not only correlation is high at the corners on the main diagonal of the map (when both $H_1$ and $H_2$ are close to nucleation or when they are both close to saturation) but also at the two other off-diagonal corners (when $H_1$ is close to nucleation and $H_2$ is close to saturation, and *vice versa*). The F magnetic domain patterns at all the four corners of the map all resemble the one pattern imprinted in the AF layer during the cooling.

**Dependence with cooling field magnitude.** The dependence of MDM with cooling field $H_{CF}$ is shown in Fig. 4. Figure 4a–d displays correlation maps measured on the ascending branch after one cycle for various $H_{CF}$ values increasing from 0 to 3,200 Oe. Associated diagonal slices are shown in Fig. 4e–h. In ZFC condition or when the magnitude of the cooling field is moderate, that is, $H_{FC} = 0$ to $\sim 2,250$ Oe, the film exhibits high MDM over an extended region throughout the magnetization cycle, and that high MDM is persistent through field cycling. The domain reversal is then mostly driven by the template imprinted in the AF IrMn layer. In ZFC conditions, the template imprinted in the AF layer has about the same amount of up and down domains, and nearly no net magnetization (coercive pattern). When propagating, the domains in the F layer tend to retrieve that particular template, very quickly after nucleation and almost all the way to saturation. Surprisingly, FC the sample under a moderate non-zero field, thus applying a significant bias, still leads to results similar to ZFC: high MDM extends through most of the magnetization cycle and persists through field cycling. Even when applying a significant net bias with a higher cooling field up to $H_{FC} = 2,250$ Oe, the imprinted domain patterns still guide the domain topology at the coercive point.

When the cooling field is increased to higher values, that is, $H_{FC} \sim 2,500$ Oe and beyond, a dramatic loss of MDM occurs over nearly the entirety of the magnetization loop and the loss gradually progresses with field cycling. When $H_{FC} \sim 2,500$ Oe (Fig. 4c), the MDM is still strong, around 90%, on the nucleation side, but gradually decreases down to $\sim 65\%$ as the magnetization progresses. In this case, the film shows a moderately high MDM plateau extending throughout most of the magnetization cycle. The average height of this plateau decreases with field cycling, from $\sim 75\%$ after one cycle down to 55% after six cycles (Fig. 4g).

When $H_{FC} = 3,200$ Oe (Fig. 4d), the film shows very low or no memory, except at the very corners of the map where the film is either at nucleation or near saturation. At such high $H_{FC}$ values, the imprinted domain pattern differs significantly from the coercive pattern. Many domains have already collapsed, leaving only a few sparse domains in the imprinted template. Consequently, the domain reversal in the F layer is mostly random and MDM is low for most of the magnetization process. However, the remaining uncollapsed domains in the imprinted template still play the role of anchors for the nucleation process and somewhat guide the final path to saturation. This causes MDM to be relatively high (up to 80%) at the extremities of the magnetization process (nucleation and saturation) but low (down to 40%) in between these points (Fig. 4h). Therefore, field cooling with a nearly saturating field ($H_{FC} \sim 3,000$ Oe) can still induce high MDM at the extremities of the magnetization process.

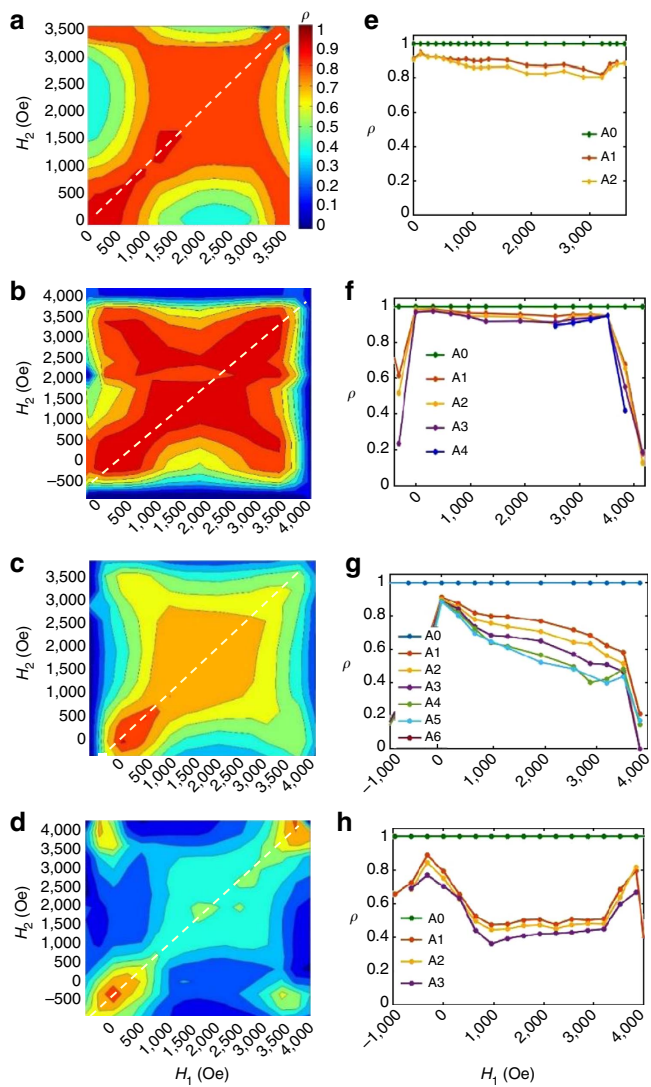

**Figure 4 | Dependence of correlation maps and associated diagonal slices with cooling field $H_{FC}$.** (**a–d**) Correlation maps $\rho(H_1, H_2)$ measured at 20 K on the ascending branch after one cycle (A1 map) under different cooling fields $H_{FC}$ (cooling starting at 400 K): (**a**) $H_{FC} = 0$ (ZFC); (**b**) $H_{FC} = 2,240$ Oe; (**c**) $H_{FC} = 2,560$ Oe; and (**d**) $H_{FC} = 3,200$ Oe. (**e–h**) Associated diagonal slices through the maps along the dotted line. Each graph includes several measured cycles (up to six cycles for $H_{FC} = 2,560$ Oe) using the notation A0, A1, A2 and so on. In this notation, the letter A stands for the ascending branch and the index corresponds to the number of cycles separating the two correlated images.

Interestingly, high correlation is also observed at the corners of lower $H_{FC}$ maps. For these lower cooling field cases, the high correlation in the central region extends flat towards the corners of the maps. The persistence of high correlation at the corners suggests the possible presence of distributions in the exchange coupling, which tend to shape the imprinted domain pattern, favouring some locations for domains to nucleate and to later vanish when approaching saturation. In other terms, exchange coupling is still the main driving force leading to magnetic memory in this material but the presence of distributions can additionally extend the high memory from the core of the magnetization cycle towards the nucleation and saturation extremities.

Finally, we attempted complementary measurements at the Mn $L_3$-edge (638 eV) in a hope to directly measure a magnetic signal from the interfacial uncompensated Mn spins in the IrMn layer and correlate it to the signal produced by the Co spins in the [Co/Pd] multilayer at the Co edge. Unfortunately, no signal could be observed at the Mn edge. Indeed, the amount of uncompensated Mn spins in the IrMn layer (presumably only a few percent of the total amount of Mn in the film[15,35]) is too small to produce a visible magnetic signal in a reasonable time frame given the sensitivity of the detector.

## Discussion

Exchange-biased [Co/Pd]/IrMn perpendicular magnetic films exhibit various degrees of MDM depending on FC conditions. The visuals in Fig. 5 illustrate this effect by showing typical magnetic domain topologies in the F and AF layers at different stages of the magnetization process under different cooling conditions. These sketches of the magnetic domains are based on actual MFM images (Fig. 5a,b). These MFM images provide a view of the magnetic domains in the superficial Co/Pd layer, at room temperature. The CXRMS measurements allow exploring the magnetic domains throughout the whole film and at low temperature, where exchange coupling is present. Our correlation results show that the amount of MDM in the F layer is ultimately set by the type of magnetic pattern imprinted in AF layer during the cooling. If the film is cooled under zero field (ZFC) or a moderate field well within the magnetization range, a maze-like domain pattern gets imprinted into the AF layer, filling most of the space (Fig. 5c,d). In this case, the F layer domain pattern will tend to replicate the imprinted pattern and therefore exhibit a very high MDM memory throughout almost the entire magnetization process, which will persist through field cycling. If the film is cooled under near-saturating or near-nucleating conditions, a sparse domain pattern gets imprinted in the AF layer (Fig. 5e,f). In this case, the F layer will only show moderately high MDM at nucleation and saturation points, where domains may be pinned down to the location of imprinted AF domains. Outside these specific points, the propagation of the domains to fill the space between the pinned sites will occur mostly randomly, leading to low MDM.

In conclusion, the behaviour of MDM in exchanged-biased F films can be linked to the amount of net exchange bias induced during the cooling process. If the film is zero-field cooled or cooled under moderate field values ($H_{FC} = 0$–2,250 Oe), that is, if the net exchange bias is small or zero, the film exhibits a very high MDM, reaching 100% extending over almost the entirety of the magnetization cycle. Moreover, MDM is persistent through field cycling. If, on the contrary, the film is cooled under higher field values, nearly saturating ($H_{FC} = 2,500$–3,200 Oe), that is, the net exchange bias is maximum, the film then gradually loses MDM throughout most of the magnetization cycle, except at nucleation and at saturation. It is expected that if the cooling field is increased to much higher values, MDM would eventually disappear completely on the entirety of the magnetization loop, even at nucleation and saturation. Partial MDM may potentially occur due to local variations in the exchange coupling or due to structural defects. We however find that strong MDM can be induced even in a relatively smooth F layer by incorporating exchange coupling with an AF template. More importantly, MDM may be controlled in specific regions of the magnetization loop by tuning the amount of net exchange bias. Applying no bias or a small bias leads to high, persistent MDM throughout the entirety of the magnetization loop. Applying a large bias result in loss of memory in the central part of the magnetization process, while keeping relatively high memory at nucleation and saturation, if the cooling field is finely tuned to near saturation.

## Methods

**Sample preparation.** The $[[Co(4\,\text{Å})/Pd(7\,\text{Å})]_{\times 12}/IrMn(24\,\text{Å})]_{\times 4}$ multilayers were deposited onto Si$_3$N$_4$-coated Si substrates via dc magnetron sputtering at room temperature under 3 mTorr Ar sputtering pressure. The thicknesses of the Co, Pd and IrMn layers and number of repeats were optimized to enhance the amount of perpendicular exchange coupling throughout the film and adjust the blocking temperature $T_B$ to be around 300 K. The Si substrates supporting low-stress 100-nm-thick Si$_3$N$_4$ membranes, of typical $2.5 \times 2.5\,\text{mm}^2$ size, had a $1\times 1\,mm^2$ square opening at their centre, to allow measurements with X-rays in transmission geometry. A circular pinhole of 20 μm in diameter was mounted on top of the film and positioned within the $1\times 1\,mm^2$ square opening. In the scattering experiment, the film-pinhole element was inserted with the pinhole facing the upstream X-ray beam, so as to achieve nearly coherent illumination of the film.

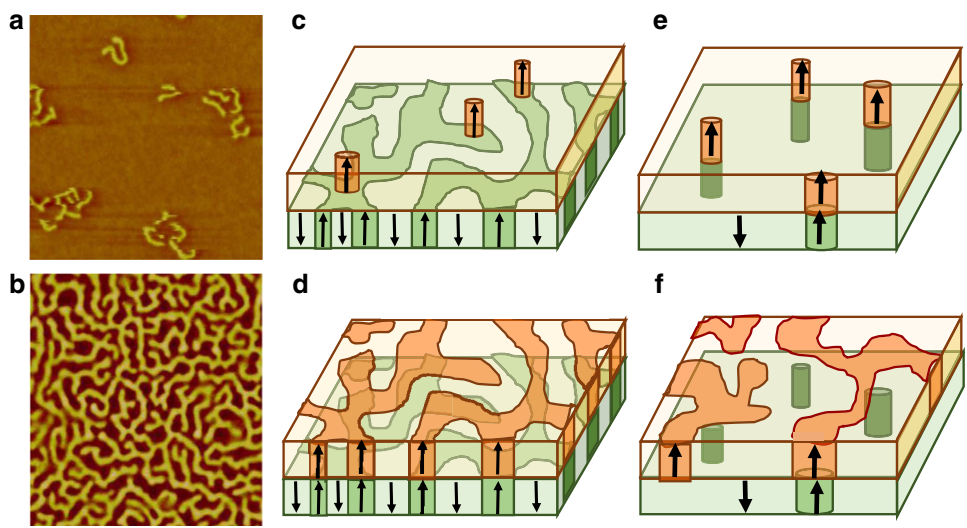

**Figure 5 | Magnetic domain topology in the [Co/Pd]/IrMn film at nucleation and in the coercive region under various FC conditions.** (**a,b**) MFM ($5 \times 5\,\mu\text{m}^2$) images taken at room temperature showing the magnetic domains in the F Co/Pd layer (**a**) at the nucleation stage and (**b**) in the coercive region ($M \sim 0$). (**c,d**) Sketches of the domain configuration for ZFC and moderate field-cooled states (**c**) at nucleation/saturation and (**d**) at the coercive point. (**e,f**) Sketches of the domain configuration in near-saturating field-cooled state (**e**) at nucleation/saturation and (**f**) at the coercive point. The imprinted pattern in the AF layer is shown in green, the domain pattern in the F layer is shown in orange. Arrows indicate magnetization direction.

**Coherent X-ray magnetic scattering experiment.** Our CXRMS measurements were carried out at the Advanced Light Source synchrotron, at the coherent soft X-ray beamline BL12.0.2 (ref. 30). The energy of the light was tuned to the magnetic resonance at the Co $L_3$-edge to optimize the magneto-optical contrast[36–38]. These CXRMS measurements were conducted with linearly polarized light. The polarization of the X-ray light produced by the U8 undulator was transverse linear in the horizontal plane ($\sigma$-polarization). The sample was mounted in transmission geometry so that its magnetization vector **M** was parallel to the direction **k** of the incident light and perpendicular to the polarization. In this configuration, the resulting magnetic scattering signal occurs in the $\sigma-\pi$ channel only (where $\pi$ denotes the transverse vertical polarization) and its amplitude is directly proportional to the dot product **k.M**[36]. The charge scattering signal, on the other hand, remains in the $\sigma-\sigma$ channel and does not interfere with the magnetic scattering. The resulting overall scattering is a simple superposition of the pure charge- and magnetic-scattering intensities. When saturating the magnetic film, the magnetic scattering component gets suppressed, leaving only the charge-scattering intensity. This reference signal is later subtracted from all the other scattering images collected throughout the magnetization loop before extracting the magnetic speckle signal. However, we note that, in our films, the structures are very uniform laterally; thus, there is very little charge scattering and any charge scattering occurs at a very different $q$ range compared with the magnetic scattering. Regarding coherence properties, the X-ray beam was spatially filtered by a 20-μm pinhole to improve its transverse coherence and limit the area of the sample probed. Over the illuminated area, about a 20-μm-diameter disk, the coherence of the X-ray beam is partial. At this energy (778 eV), the transverse coherence length of the X-ray beam is $\sim 5\,\mu m^{30}$. We have estimated that the associated degree of coherence[29] in our speckle patterns is $\sim 15-20\%$. This partial coherence proves not to be a limiting factor in our cross-correlation work. In our analysis, we separate the purely coherent signal from the incoherent signal before the correlations, by using an iterative smoothing technique[32]. The smoothed scattering patterns, which essentially represents the incoherent part of the scattering signal, are subtracted from the collected scattering images so as to only leave the purely coherent part of the scattering signal (speckle). We then perform our cross-correlations directly on the speckle patterns. This approach enables us to evaluate MDM with a good accuracy and no information is lost due to the partial coherence of the light.

**Speckle cross-correlation technique.** The amount of correlation between two speckle images is evaluated through a normalized correlation coefficient defined as follows:

$$\rho = \frac{\sum_{i,j}(A\times B)_{i,j}}{\sqrt{\sum_{i,j}(A\times A)_{i,j}\sum_{i,j}(B\times B)_{i,j}}}$$

where $A$ and $B$ represent the two speckle patterns to be correlated and $\times$ represents the correlation operation[32]. The correlation patterns $A\times B$, $A\times A$ and $B\times B$ are two-dimensional images of same size than images $A$ and $B$ (an example of such correlation pattern is shown in Fig. 1i, a close-up). Correlations patterns usually exhibit a peak whose width corresponds to the average speckle size and whose intensity reflects the amount of correlation between the speckle patterns. The indices $(i,j)$ refer to pixel positions in these correlation patterns, so that the summation $\sum$ allows integrating the signal under the correlation peak. This integration is performed both on the cross-correlated signal $A\times B$ and the auto-correlated signals $A\times A$ and $B\times B$ for normalization purposes. The resulting coefficient $\rho$ lies between 0 and 100%. The maximum value is reached when the two speckle patterns $A$ and $B$, and consequently their associated magnetic domain patterns, are nearly identical.

### Acknowledgements
This work, including the VSM and MFM measurements, as well as the collection and analysis of the synchrotron scattering data, was supported by internal ORCA and MEG grants at BYU. Correlation work was supported by time on supercomputer Marylou at BYU. Work at UCSD on sample fabrication and characterization, as well as discussion of the results was supported by the research programmes of the US Department of Energy (DOE), office of Basic Energy Sciences (Award Number DE-SC0003678). The scattering data were collected at the Advance Light Source, a US DOE Office of Science User Facility operated by the Lawrence Berkeley National Laboratory. We thank Steve Kevan for facility support with the coherent scattering chamber at the ALS.

### Author contributions
This study was conceived by K.C. and E.E.F. E.E.F. synthesized and characterized the magnetic thin films. K.C. conducted the study, collected the coherent X-ray scattering images at the ALS, the magnetometry data and the MFM images at BYU. A.S. and M.R. analysed the scattering images and computed the speckle correlation maps. K.C. gathered all the maps and slices, and interpreted the results. E.E.F. provided feedback on the results. K.C. wrote the paper with feedback from all the authors.

### Additional information
**Competing financial interests:** The authors declare no competing financial interests.

**Reprints and permission** information is available online at http://npg.nature.com/reprintsandpermissions/

**How to cite this article:** Chesnel, K. *et al.* Shaping nanoscale magnetic domain memory in exchange-coupled ferromagnets by field cooling. *Nat. Commun.* 7:11648 doi: 10.1038/ncomms11648 (2016).

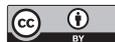